\documentclass[review]{elsarticle}
\usepackage{lineno,hyperref}
\usepackage[english]{babel}
\usepackage[numbers]{natbib}
\usepackage{xcolor}
\usepackage{latexsym,amsmath,amssymb,amsbsy,graphicx,geometry}
\modulolinenumbers[5]

\begin{document}
	
	\begin{frontmatter}

\title{A Unified Empirical Equation for Determining the Mechanical Properties of Porous NiTi Alloy: From Nanoporosity to Microporosity}

\author[kfu]{B.N. Galimzyanov}

\author[kfu]{G.A. Nikiforov}

\author[tgu]{S.G. Anikeev}

\author[tgu]{N.V. Artyukhova}

\author[kfu]{A.V. Mokshin}
\cortext[cor1]{Corresponding author}
\ead{anatolii.mokshin@mail.ru}

\address[kfu]{Kazan Federal University, 420008 Kazan, Russia}
\address[tgu]{Tomsk State University, 634050 Tomsk, Russia}

\begin{abstract}
	The mechanical characteristics of a monolithic (non-porous) crystalline or amorphous material are described by a well-defined set of quantities. It is possible to change the mechanical properties by introducing porosity into this material: as a rule, the strength values decrease with the introduction of porosity. Thus, porosity can be considered as an additional degree of freedom that can be used to influence the hardness, strength and plasticity of the material. In the present work, using porous crystalline NiTi as an example, it is shown that the mechanical characteristics such as the Young’s modulus, the yield strength, the ultimate tensile strength, etc. demonstrate a pronounced dependence on the average linear size $\bar{l}$ of the pores. For the first time, an empirical equation is proposed that correctly reproduces the dependence of the mechanical characteristics on the porosity $\phi$ and on the average linear size $\bar{l}$ of the pores in a wide range of sizes: from nano-sized pores to pores of a few hundred microns in size. This equation correctly takes into account the limit case corresponding to the monolithic material. The obtained results can be used directly to solve applied problems associated with the design of materials with the necessary combination of physical and mechanical characteristics, in particular, porous metallic biomaterials.
\end{abstract}

\begin{keyword}
NiTi; porous NiTi; mechanical properties; porous structure
\end{keyword}

\end{frontmatter} 

\section{Introduction}

Porous materials have a unique combination of physical and mechanical properties that is not found in their monolithic analogues~\cite{Thomas_2020,Yan_Xiao_2021,Orellano_Sanz_2022,Zadpoor_2019,Anikeev_Kaftaranova_2023}. For example, the presence of percolating voids and a significant specific surface area typical of porous materials provide their lower density, lower thermal conductivity and high catalytic activity compared to non-porous analogues~\cite{Liu_2014}. The potential functional applicability of porous materials is largely determined by their mechanical properties $\mathcal{M}$~\cite{Kruzic_2016,Monogenov_2022,Brothers_2006,Galimzyanov_Nikiforov_2020,Filipe_Loura_2022}. Here the quantity $\mathcal{M}$ denotes a set of properties such as the Young's modulus $E$, the yield stress $\sigma_{y}$, the shear modulus $K_{s}$, the bulk modulus $G_{b}$, etc., i.e.
\begin{equation}\label{eq_mechprop}
\mathcal{M}=\{E,\,\sigma_{y},\,K_{s},\,G_{b},\,\dots\}.
\end{equation}
It is well known that the mechanical properties $\mathcal{M}$ of the monolithic crystalline or amorphous material are mainly determined by thermodynamic conditions: the pressure $p$ and the temperature $T$~\cite{Wu_2014,Elkenany_2019,Galimzyanov_JNCS_2021,Galimzyanov_Mokshin_JNCS_2021}.

Let us assume that there is a bulk porous material whose characteristic dimensions -- length, width and height -- are comparable. In this case, the linear size of the material can be estimated as $\sqrt[3]{V}$, where $V$ is the volume of the material. Obviously, such specific porous materials as porous rods, membranes, porous thin films are not included in this category. The porosity of the bulk material can be \textit{quite unambiguously} characterized by two quantities: the pore size distribution function $P(l)$ and the interpore partition size distribution function $P(g)$. Other characteristics of the porous structure -- such as the most probably pore shape, the type of pores, the average pore surface area, etc. -- should be directly or implicitly related to $P(l)$ and $P(g)$. On the other hand, it is convenient to characterize the porous structure in terms of scalar quantities. Therefore, the functions $P(l)$ and $P(g)$ can be replaced by the pair of quantities with some \textit{loss of uniqueness} characterization:
\begin{equation}\label{eq_averagequant}
\bar{l}=\frac{\int l\,P(l)\,dl}{\int P(l)\,dl} \hskip 1cm \text{and} \hskip 1cm \bar{g}=\frac{\int g\, P(g)\,dg}{\int P(g)\,dg},
\end{equation}
where $\bar{l}$ is the average linear size of the pores, $\bar{g}$ is the average linear size of the interpore partitions. In practice, both $\bar{l}$ and $\bar{g}$ can be estimated by statistical analysis of ``patterns'' (pore contours) on thin sections of the material. In addition, these two quantities correlate with a well-known quantity such as the total porosity of the material $\phi=V_{0}/V$, where $V_{0}$ is the volume of empty space in the material, the total volume of which is $V$. The porosity $\phi$ is dimensionless and $\phi\in (0;\,1)$; for $\phi=0$ we have the case of the monolithic material. It is therefore clear that it is convenient to use the porosity $\phi$ instead of one of the two quantities $\bar{l}$ and $\bar{g}$ to characterize a porous material~\cite{Ji_S_2006,Honicke_2018,GBN_2021,Liu_Pan_2022}.

\begin{figure}[ht!]
	\centering
	\includegraphics[width=9cm]{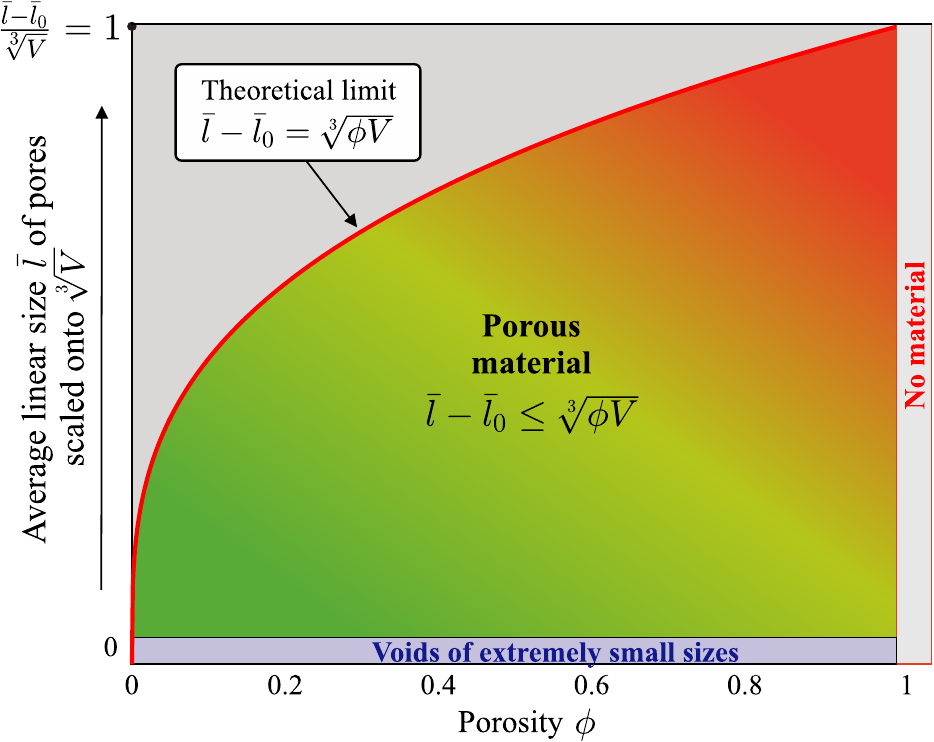}
	\caption{Diagram demonstrating the possible range of $\bar{l}$ and $\phi$ for which a porous material is realized. Here, the average linear size $\bar{l}$ of the pores is measured in the reduced unit $\sqrt[3]{V}$ and takes values in the range $(\bar{l}-\bar{l}_{0})/\sqrt[3]{V}\in(0;\,1)$; $V$ is the volume of the material. The thick red line indicates the theoretical limit, which results from equation $\bar{l}-\bar{l}_{0}=\sqrt[3]{\phi V}$. A porous material can only be realized in the coloured region, where the condition $\bar{l}-\bar{l}_{0}\leq\sqrt[3]{\phi V}$ is satisfied. Limit values $\phi=0$ and $(\bar{l}-\bar{l}_{0})/\sqrt[3]{V}=0$ correspond to the monolithic material, while $\phi=1$ and $(\bar{l}-\bar{l}_{0})/\sqrt[3]{V}=1$ material is not realized.}\label{fig_1}
\end{figure}

In fact, one has two physical quantities: (i) the average linear size of a pore and (ii) the average linear pore size for a whole system. First, the average linear size of a pore is defined as the average of all possible distances between opposite walls of a considered single pore. The distance between opposite walls means the length of the segment that is common to the perpendiculars drawn from the opposite walls. Second, the average linear pore size $\bar{l}$ for a whole system is just an average characteristic for all the pores of the material. It can be simply defined as the normalized first moment of the pore size distribution function $P(\bar{l})$ [see Eq.~(\ref{eq_averagequant})]. In the general case, we have
\[
\bar{l}\in(0;\,\sqrt[3]{V}) \hspace{0.3cm}\text{or} \hspace{0.3cm} \bar{l}\in[\bar{l}_{0};\,\sqrt[3]{V}).
\]
Here, $\bar{l}_{0}$ is the quantity characterising the average interatomic distance in the monolithic material. The value of this quantity can approximate the size of point defects. If we define the average size of the atoms forming the alloy as $d$, then the value of $\bar{l}_{0}$ can be estimated as $\bar{l}_{0}\simeq(1/3)d$. For $\bar{l}=\bar{l}_{0}$ we have the monolithic material; and for $\bar{l}\to\sqrt[3]{V}$ there is no material as such. The range of possible values of $\bar{l}$ is limited by the value of the porosity $\phi$ according to the following inequality:
\begin{equation}\label{eq_philimit}
\bar{l}-\bar{l}_{0}\leq\sqrt[3]{\phi V}.
\end{equation}
On the ($\bar{l}$, $\phi$)-diagram, a porous material can only be realized within a strictly defined range of values of these quantities, as shown in Figure~\ref{fig_1}.

In the case of a porous material with non-zero values of $\bar{l}$ and $\phi$, an arbitrary mechanical characteristic from the set $\mathcal{M}$ [see relation~(\ref{eq_mechprop})] will be a function of $\bar {l}$ and $\phi$, i.e.
\begin{equation}\label{eq_Mfunc}
\mathcal{M}=\mathcal{M}(\bar{l},\,\phi).
\end{equation}
Almost all strength properties of the material decrease with increasing $\bar{l}$ and $\phi$~\cite{Kolesnikova_2017,Torres_Mushref_2017,Su_Rao_2020,Galimzyanov_Mokshin_JPCM_2022}. In general, the following conditions can be formulated
\begin{subequations}\label{eq_cond}
	\begin{equation}
	\frac{\partial \mathcal{M}(\bar{l},\,\phi)}{\partial\phi}\leq 0,\label{eq_a}
	\end{equation}
	\begin{equation}
	\frac{\partial \mathcal{M}(\bar{l},\,\phi)}{\partial\bar{l}}\leq 0, \label{eq_b}
	\end{equation}
\end{subequations}
which characterize the decreasing strength properties with increasing porosity and/or pore size.

The condition (\ref{eq_a}) is taken into account by almost all known models. So, the example of such a model is the Bert's power law equation for porous materials with elliptically shaped pores~\cite{Bert_1985,Choren_2013,Ji_Gu_2006}:
\begin{equation}\label{Bert_model}
\mathcal{M}(\phi)=\mathcal{M}_{0}\left(1-\frac{\phi}{\phi_{m}}\right)^{K\phi_{m}},
\end{equation}
where $K=0.75+1.25(b/c)$. Here, the parameter $c$ is the average linear size of the pores along the applied stress; $b$ is the average linear size of the pores perpendicular to the direction of the applied stress; $\mathcal{M}_{0}$ is the mechanical properties of the monolithic material; $\phi_{m}$ is the maximum possible porosity for the material. Equation (\ref{Bert_model}) is a modification of the linear Rossi expression~\cite{Rossi_1968}, and this expression is mainly used to describe the dependence of elastic properties on porosity. In Eq.~(\ref{Bert_model}), the degree function is required for a more accurate description of the mechanical properties at porosity $\phi>0.2$. In addition, in the case of an isotropic porous system with isolated pores whose sizes have a random distribution, the dependence of mechanical properties on porosity can be described by Mackenzie's semi-empirical equation~\cite{Mackenzie_1950}:
\begin{equation}\label{Mackenzie_model}
\mathcal{M}(\phi)=\mathcal{M}_{0}\left(1-d\phi+g\phi^{2}\right).
\end{equation}
Here, the parameters $d$ and $g$ characterize the shape of the closed pores. At the same time, Eq.~(\ref{Mackenzie_model}) incorrectly describes the quantity $\mathcal{M}(\phi)$ in the case of a porous system with open pores, due to the difficulties associated with the determination of the linear dimensions of the pores. There are empirical model that do not explicitly take into account the morphology of a porous system. The parameters in such models are fitted through approximation of empirical data for a sample with a specific synthesis protocol, which makes the application of such models more universal. An example of such a model is the Ryshkewitch-Duckwoth exponential equation~\cite{Ryshkewitch_1953, Duckworth_1953}:
\begin{equation}\label{eq_Duckworth}
\mathcal{M}(\phi)=\mathcal{M}_0\exp(-B\phi),
\end{equation}
where the exponent $B$ is the empirical constant. It is important to note that this equation correctly describes the behaviour of mechanical properties at porosity values up to $0.6$. Bal'shin power law equation is suitable for reproducing the mechanical properties of porous materials over a wide range of porosities~\cite{Balshin_1949}:
\begin{equation}\label{eq_Balshin}
\mathcal{M}(\phi) = \mathcal{M}_0(1-\phi)^{n},
\end{equation}
where the exponent $n$ is the parameter that depends on the type of material and on the deformation protocol of the material. Thus, all the above equations take into account the fact that the introduction of porosity into the material leads to a significant decrease in the elastic, plastic and strength properties according to the exponential or power law.

Let the porosity $\phi$ be fixed. The average linear size $\bar{l}$ of the pores can vary over a fairly wide range, and the maximum possible pore size $\bar{l}_{\mathrm{max}}$ cannot exceed the size of the material. For such a fixed porosity $\phi$, the maximum possible pore size $\bar{l}_{\mathrm{max}}$ can be realized in a material with a single large pore; and a porous material with the minimum possible pore size $\bar{l}_{\mathrm{min}}$ is manifested by the presence of a certain number of pores with an extremely small non-zero size, i.e. $\bar{l}_{\mathrm{min}}\neq0$. The aim of the present work is to find out how the mechanical strength properties of the porous material depend on the pore size. It is assumed that a fixed porosity ($\phi=$const) can be provided by both millimeter-sized pores and extremely small pores of the order of tens of nanometres. This problem is solved using the example of such a well-known metal alloy as NiTi under normal conditions (temperature $T=300$\,K and pressure $p=1$\,atm).

The NiTi alloy is considered as a promising material for use in engineering and medicine due to its unique combination of physical properties, including shape memory effect, superelasticity, high corrosion resistance and biocompatibility~\cite{GBN_2021,Greiner_Oppenheimer_2005,Anikeev_Artyukhova_2022}. Materials based on this alloy have found wide application as actuators in microelectromechanical devices~\cite{Stachiv_I_2021}, in the aerospace industry~\cite{Hartl_D_2007,Costanza_G_2020}, in the manufacture of medical implants~\cite{Kapoor_D_2017,Anikeev_Garin_2018}. Porous NiTi has a large specific surface area, which allows it to be used to treat or replace damaged human organs due to its ability to penetrate and implant biological tissue into the pore space, where this tissue develops naturally~\cite{Aihara_2019}. At the same time, porous NiTi, like most other porous materials, has a lower resistance to deformation than its monolithic analogue~\cite{Ji_S_2006}. It should be noted that the introduction of porosity into NiTi alloy can change its microstructure limiting phase transitions associated with austenitic-martensitic transitions~\cite{Anikeev_Artyukhova_2022}. The presence of defects and free surface due to porosity as well as the formation of a percolating branched structure by the crystalline matrix can limit the superelasticity and shape memory effect~\cite{Liu_Wu_2009,Hosseini_Sadrnezhaad_2009}. In the present work, we do not study the effect of porosity on the austenitic-martensitic phase transitions because the porosity of the considered porous NiTi samples is fixed.

The NiTi alloy is an intermetallic compound whose crystal structure has been identified as the high-temperature B2 cubic phase (austenite). The NiTi alloy can undergo a martensitic transition to the low-temperature orthorhombic B19' structure (martensite) with monoclinic distortion through an intermediate rhombohedral R-phase. This transition can be induced by heat treatment and forces such as tension, compression and bending. The structural transition occurs on cooling through the scheme B2$\rightarrow $R$\rightarrow$B19', while the reverse martensitic transition B19'$\rightarrow $R$\rightarrow$B2 is realized on heating~\cite{Aihara_2019}. Due to the austenitic-martensitic phase transitions, NiTi alloy has a shape memory effect and superelasticity - the ability to fully recover its original shape after removal of an external load, which can significantly exceed the yield strength of the material~\cite{Sadiq_Wong_2010,R_Casati_2017}. At normal atmospheric pressure, NiTi has the relatively high liquidus temperature $T_{l}\simeq1580$\,K, which allows this alloy to be classified as heat resistant~\cite{Gao_Bodunde_2022}. Therefore, NiTi is widely used in the aerospace, automotive, electronics and biomedical industries, for example, in the manufacture of self-sealing thermomechanical joints, thermally sensitive mechanical actuators, robotic elements, electronic device actuators, instruments and implants for cranio-cerebral and maxillofacial surgery, traumatology, orthopaedics, etc.~\cite{Zadpoor_2019,Huang_Ford_2020}.

\section{Materials and Methods}
{\tiny }

To solve the problem formulated above, we propose to consider monolithic NiTi alloy without pores as well as porous NiTi with micron-sized pores prepared by powder sintering method [Figure~\ref{fig_2}(a)] and nanoporous NiTi obtained through simulation [Figure~\ref{fig_2}(b)]. The following conditions are considered here:

\begin{figure*}[ht!]
	\centering
	\includegraphics[width=13cm]{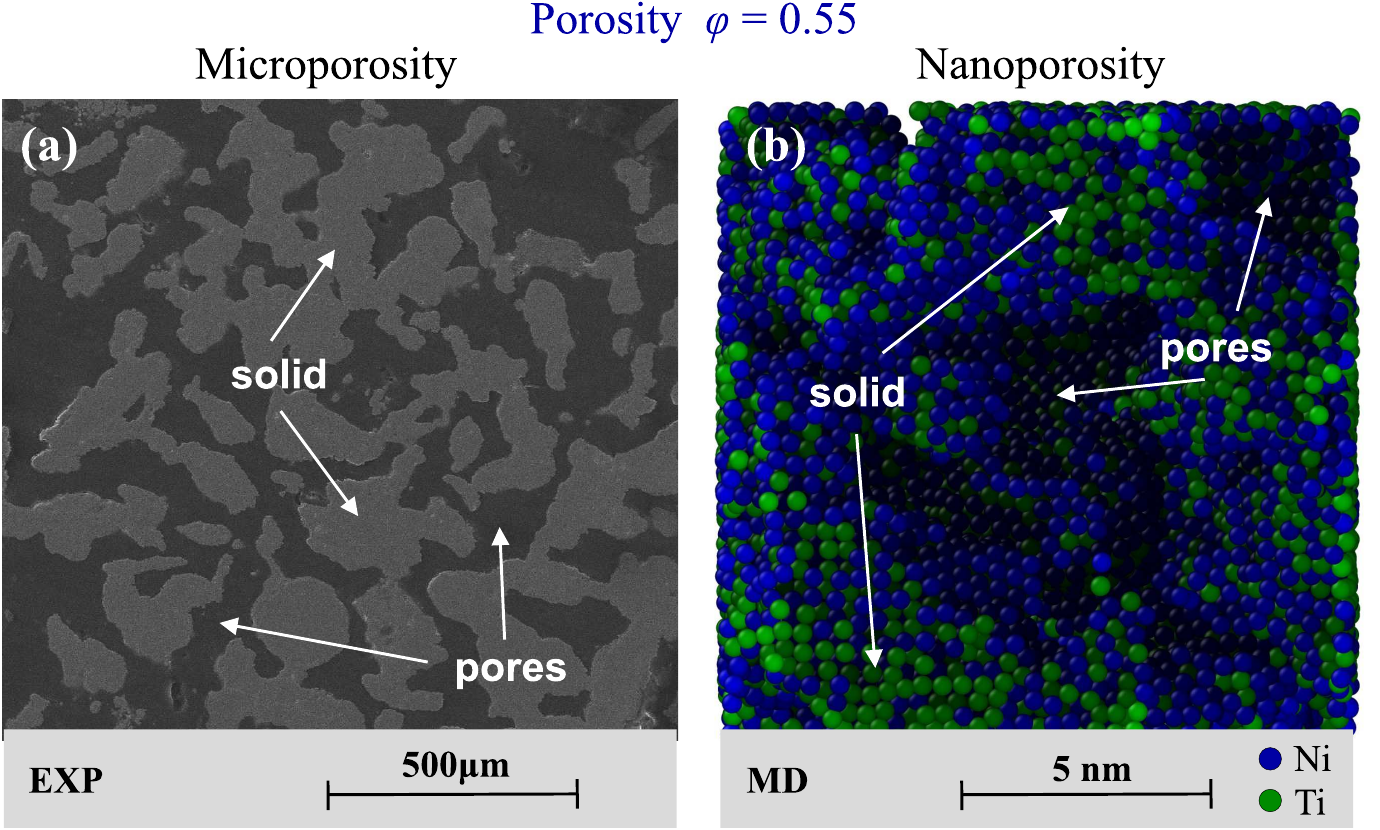}
	\caption{(a) Thin section of porous NiTi with micron-sized pores obtained by powder sintering. The pores and the solid matrix are identified by scanning electron microscopy through analysis of the thin sections of porous samples, where the pores are black because they are filled with impregnating resin and the solid frame is grey in color. (b) Snapshot of nanoporous NiTi obtained by simulation using OVITO software~\cite{Stukowski_2010}. The solid matrix and the pores are indicated by arrows.}\label{fig_2}
\end{figure*}

(I) \underline{Monolithic material}. This case can be considered as the limit case of a porous material, where the average linear pore size $\bar{l}$ approximates the linear sizes of the point defects, i.e. $\bar{l}\simeq\bar{l}_{0}\simeq0.09$\,nm.
The mechanical properties of monolithic NiTi depend on the type of crystal lattice. For example, for B2 phase monolithic NiTi the Young's modulus $E$, which characterizes the ability of the material to resist tensile stress under elastic deformation, is $E\simeq75\pm10$~GPa, whereas  $E$ for the B19' phase is $E\simeq32\pm9$~GPa [see Table~\ref{tab_1}]. The yield strength, which determines the limit of elastic behavior and the onset of plastic deformation on the stress-strain curve, is $\sigma_{y}\simeq0.8\pm0.1$~GPa for austenite and $\sigma_{y}\simeq0.1\pm0.03$~GPa for martensite. The ultimate strength, which is the stress at which the material begins to fracture, is $\sigma_{f}\simeq1.9\pm0.3$~GPa~\cite{B_AL_Mangour_2013,S_Hellberg_2020,A_Jalaeefar_2013}. The ultimate strain $\epsilon_{f}$, which corresponds to ultimate strength, is $\epsilon_{f}>15$~\% and depends on the quality of the raw material as well as on the production protocol of the crystalline alloy (i.e. on the degree of mixing of the alloy components, on the heating temperature, on the cooling rate of the melt).

\begin{figure}[ht!]
	\centering
	\includegraphics[width=9cm]{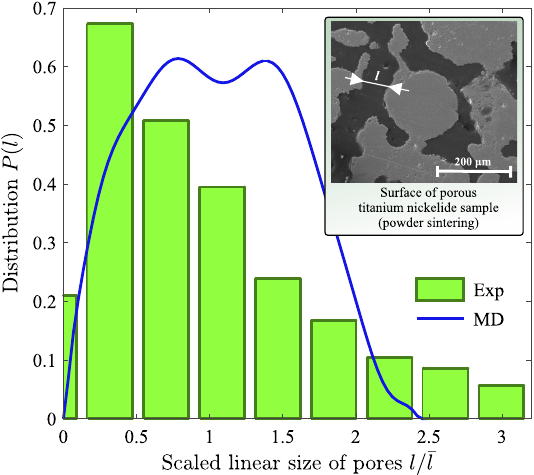}
	\caption{Pore distributions $P(l)$ over linear sizes obtained for nanoporous NiTi (through simulation) and for porous NiTi with micron-sized pores (by experiment) with the porosity $\phi=0.55$ at the temperature $T=300\,$K. Inset: image of the thin section obtained by powder sintering, where interpore partitions (light areas) and voids (dark areas) are shown. The linear pore size $l$ is represented as the reduced quantity $l/\bar{l}$, where $\bar{l}$ is the average linear size of the pores.}\label{fig_3}
\end{figure}

\begin{figure*}[ht]
\centering
	\includegraphics[width=15.0cm]{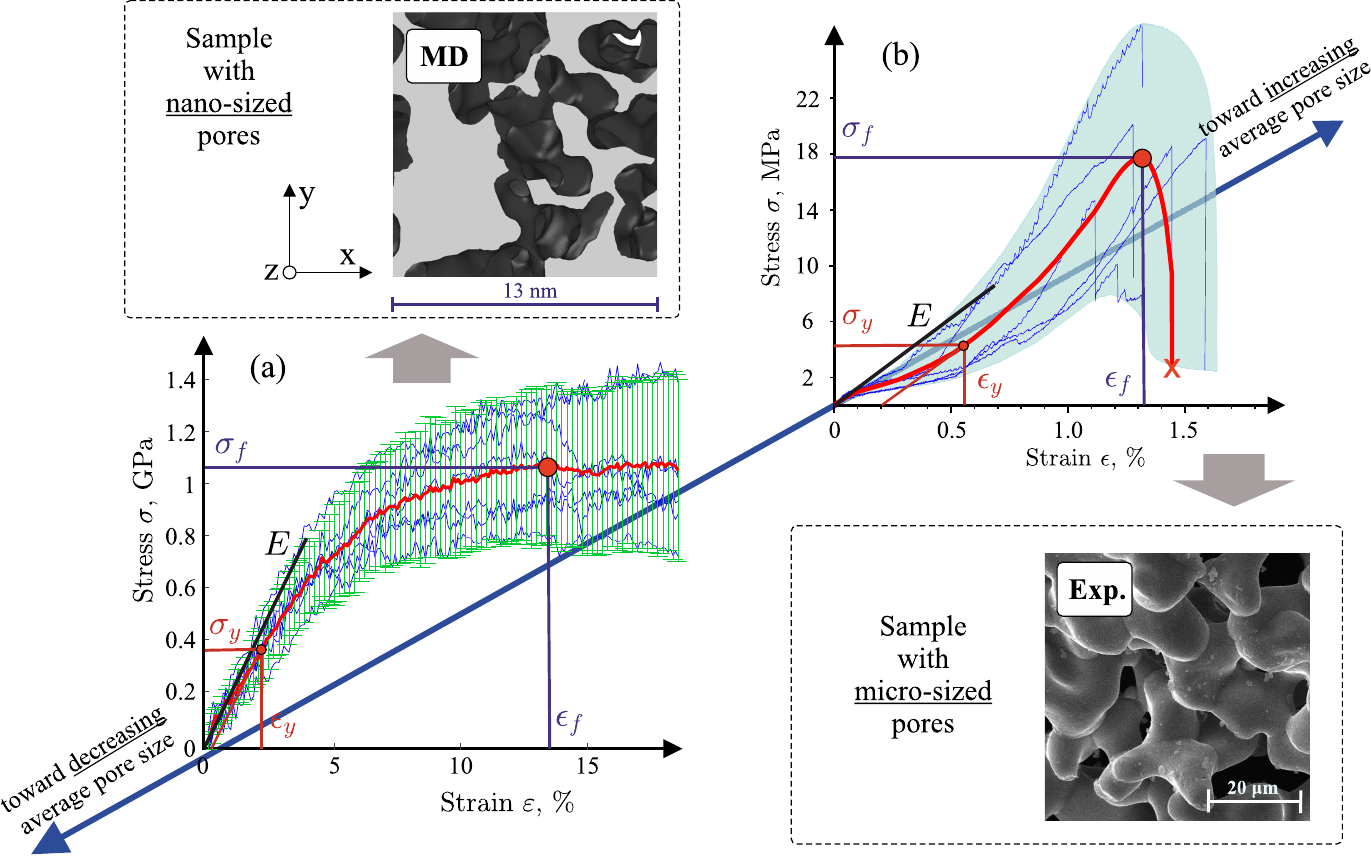}
	\caption{Stress-strain curves of porous NiTi under uniaxial tension obtained for samples with (a) nano-sized pores and (b) micron-sized pores by means of non-equilibrium molecular dynamics (MD) simulations and experimental studies (Exp.). Insets: thin sections of nanoporous NiTi and porous NiTi with the characteristic pore size $\bar{l}\simeq78$~$\mu$m, obtained from MD simulation and from experiment, respectively. The red solid curves represent the averaged results.}\label{fig_4}
\end{figure*}

(II) \underline{Nanoporous alloy} ($\phi=0.55$, $\bar{l}\simeq4.4$\,nm).
According to the classical definition~\cite{Lu_Zhao_2004}, the nanoporous material is a porous material with characteristic pore sizes of less than $100$ nm. In this case, we propose to consider NiTi with the porosity $\phi=0.55$, where the average linear pore size is $\bar{l}\simeq4.4$\,nm. This value of the quantity $\bar{l}$ follows from the pore distribution $P(l)$ over linear sizes obtained by molecular dynamics simulations of the nanoporous alloy and shown in Figure~\ref{fig_3}. The obtained distribution $P(l)$ is the result of averaging over five independent molecular dynamics simulations performed from different initial configurations under identical thermodynamic conditions. Non-equilibrium molecular dynamics simulations of this nanoporous alloy have been used to determine stress-strain curves and to identify key mechanical properties. Figure~\ref{fig_4}(a) shows the stress-strain curves for uniaxial tensile strain obtained for different porous systems from independent molecular dynamics calculations. These stress-strain curves vary widely due to the inhomogeneous structure of the porous systems. Therefore, these curves are averaged and the most probable stress-strain curve is found [in Figure~\ref{fig_4}(a) this curve is shown in red]. The Young's modulus $E$ is determined as the slope of the linear part of this curve in the low strain region corresponding to the elastic regime. The yield strength (or conventional yield strength) $\sigma_{y}$ is defined as the stress corresponding to $0.2$\% plastic deformation. The ultimate strength $\sigma_{f}$ is the maximum stress on the averaged stress-strain curve. Thus, for uniaxial tensile strain, we find that nanoporous NiTi is characterized by the Young's modulus $E\simeq17\pm3$~GPa, the yield strength $\sigma_{y}\simeq0.38\pm0.15$~GPa and the ultimate strength $\sigma_{f}\simeq1.1\pm0.3$~GPa. The limit value of the deformation $\epsilon_{f}$, at which the destruction of the nanoporous alloy is initiated, is $\epsilon_{f}\sim13.4$~\%, which is comparable to the tensile limit of monolithic NiTi. Details of the molecular dynamics calculations associated with the preparation of nanoporous NiTi and the determination of mechanical properties are given in ``Appendix: Experimental and simulation details''.

\begin{table*}[tbh]
	\caption{Mechanical properties and parameters of monolithic NiTi and porous NiTi: porosity $\phi$, average linear size $\bar{l}$, Young's modulus $E$, yield strength $\sigma_y$, ultimate strength $\sigma_f$, strain at the yield strength $\epsilon_{y}$ and strain at the ultimate strength $\sigma_{f}$. The experimental values of the mechanical properties for monolithic NiTi are taken from Refs.~\cite{B_AL_Mangour_2013,S_Hellberg_2020,A_Jalaeefar_2013}, whereas the mechanical properties of porous NiTi with micron-sized pores and nano-sized pores are defined in the present study.}	
	\begin{tabular}{cccc}
		\hline
		\textbf{Properties} & \textbf{Exp (monolithic)} & \textbf{MD (nano-sized pores)} & \textbf{Exp (micron-sized pores)}  \\
		\hline
		$\phi$  		& $0$  & $0.55$ & $0.55$ \\
		$\bar{l}$,~nm 	& $\bar{l}_{0}\simeq0.09$ & $4.4\pm0.6$  & $(78\pm7)\times10^{3}$ \\
		\hline
		$E$,~GPa		& $75\pm10$  & $17\pm3$  & $5.85\pm0.8$ \\
		$\sigma_y$,~GPa	& $0.8\pm0.11$  & $0.38\pm0.15$ & $(4.2\pm3)\times10^{-3}$ \\
		$\sigma_f$,~GPa	& $1.9\pm0.3$  & $1.1\pm0.3$ & $(17.8\pm8)\times10^{-3}$ \\
		$\epsilon_{y},~\%$  & -- & $4.7$ & $0.56$ \\
		$\epsilon_{f},~\%$  & $>15$ & $13.4$ & $1.32$ \\
		\hline
	\end{tabular}\label{tab_1}
\end{table*}

(III) \underline{Porous alloy with micron-sized pores} ($\phi=0.55$, $\bar{l}\simeq78$\,$\mu$m).
In this case, we propose to consider NiTi with the porosity $\phi=0.55$ and whose characteristic pore size is $\bar{l}\simeq78$\,$\mu$m. This alloy was prepared by powder sintering (experimental details are given in the section ``Appendix: Experimental and simulation details''). The value $\bar{l}\simeq78$\,$\mu$m was obtained from the pore size distribution $P(l)$, which, in turn, was determined using the stereometric method of random section planes. The pore size distribution shown in Figure~\ref{fig_3} is the result of averaging over different thin sections. The obtained samples show the percolating porous structure, where the voids are in the form of channels of approximately the same width, mainly due to the peculiarities of sintering the mixture of powdered raw material. The mechanical properties of porous NiTi with micron-sized pores are determined from the stress-strain curve obtained by averaging the experimental results for different samples under identical thermodynamic and deformation conditions. Figure~\ref{fig_4}(b) shows that the transition region from elastic to plastic deformation is blurred at strains $<0.5$\%. This is mainly due to the fact that structural transitions between the austenitic B2-phase and the martensitic B19'-phase occur at the tensile strain, which causes the pseudoplastic behavior of the porous material at strains from $0.1$ to $0.3$\%. As a result, it was found that for porous NiTi alloys the Young's modulus is $E\simeq5.85\pm0.8$\,GPa, the yield strength is $\sigma_{y}\simeq(4.2\pm3)\times10^{-3}$\,GPa and the ultimate strength is $\sigma_{f}\simeq(17.8\pm8)\times10^{-3}$\,GPa [see Table~\ref{tab_1}]. Such relatively low strength characteristics in the case of samples with micron-sized pores can be related to the quality of the initial powdered raw material and the powder sintering conditions~\cite{G_Dharmalingam_2019}. Namely, in the case of porous NiTi, the use of powder metallurgy methods can lead to the formation of defects inherent in the crystal structure, as well as the formation of nano-sized cracks at the places of powder sintering, which significantly reduce the strength characteristics of the porous alloy~\cite{Taheri_AM_2017,Yu_Tao_Jian_2015}.

\section{Results and discussion}

Let us define the general conditions for an arbitrary mechanical property from the set $\mathcal{M}$ as a quantity depending on the average linear size $\bar{l}$ and on the porosity $\phi$. First, it must satisfy the conditions (\ref{eq_cond}) characterising that the value of $\mathcal{M}$ decreases with increasing $\bar{l}$ and $\phi$. Second, in the case of the monolithic material, we have
\begin{subequations}\label{eq_conds}
\begin{equation}\label{eq_cond_1}
\bar{l}=\bar{l}_{0}, \hspace{0.5cm} \phi=0 \hspace{0.5cm} \text{and} \hspace{0.5cm} \mathcal{M}(\bar{l},\,\phi)=\mathcal{M}_{0}.
\end{equation}
In addition, if there is only one void or an extremely small number of voids with the linear size $\bar{l}$, which is much smaller than the size of the material $\sqrt[3]{V}$, then the values of the mechanical properties will also approach the values of the monolithic material:
\begin{equation}\label{eq_cond_3}
\frac{\bar{l}-\bar{l}_{0}}{\sqrt[3]{V}}\to 0, \hspace{0.5cm} \phi\to 0 \hspace{0.5cm} \text{and} \hspace{0.5cm} \mathcal{M}(\bar{l},\,\phi)\to\mathcal{M}_{0}.
\end{equation}
Furthermore, we can formulate the following boundary condition:
\begin{equation}\label{eq_cond_4}
\bar{l}\to\bar{l}_{0}, \hspace{0.5cm} \phi\in(0;\,1) \hspace{0.5cm} \text{and} \hspace{0.5cm} \mathcal{M}(\bar{l},\,\phi)\to\mathcal{M}_{0},
\end{equation}
which corresponds to the monolithic material with a set of extremely small voids. And, finally, for the limit situation, where the material is completely absent, we have
\begin{equation}\label{eq_cond_2}
\bar{l}=\sqrt[3]{V}, \hspace{0.5cm} \phi=1 \hspace{0.5cm} \text{and} \hspace{0.5cm} \mathcal{M}(\bar{l},\,\phi)=0.
\end{equation}
\end{subequations}

\begin{figure}[ht!]
	\includegraphics[width=12cm]{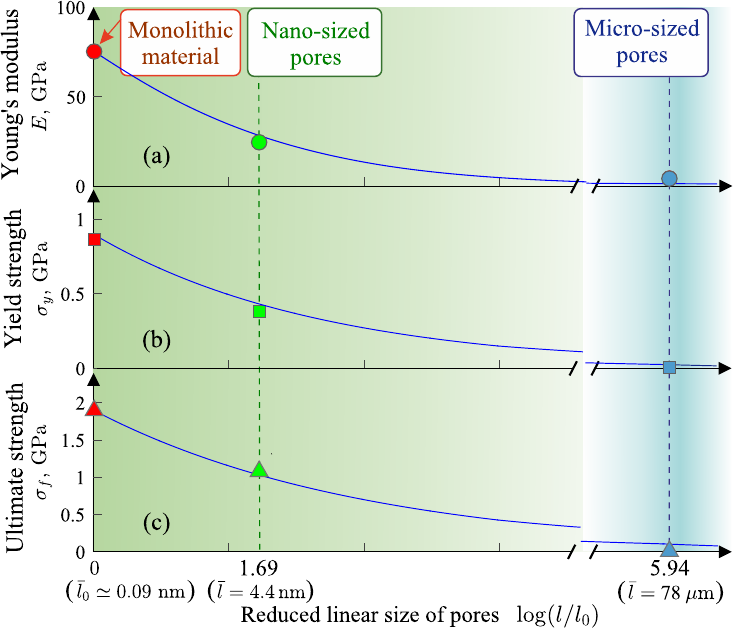}
	\caption{Dependence of the mechanical properties of NiTi on the reduced average linear pore size $\log(\bar{l}/\bar{l}_{0})$. The solid curve is the result of equation $\mathcal{M}=\mathcal{M}_{0}(\bar{l}_{0}/\bar{l})^{\alpha}$, where $\alpha\simeq0.43$ for the Young's modulus $E$ and $\alpha\simeq0.22$ for the yield strength $\sigma_{y}$ and the ultimate strength $\sigma_{f}$.}\label{fig_5}
\end{figure}

For the $\phi$-dependence of an arbitrary mechanical characteristic from the set $\mathcal{M}$, we propose to use the empirical Bal'shin equation (\ref{eq_Balshin}), which ensures the fulfillment of the conditions (\ref{eq_cond_1}) and (\ref{eq_cond_2}). On the other hand, we obtain that the $\bar{l}$-dependence of an arbitrary mechanical characteristic is reproduced by the power law of the form
\begin{equation}\label{eq_M_1}
\mathcal{M}(\bar{l})\propto\left(\frac{\bar{l}_{0}}{\bar{l}}\right)^{\alpha},
\end{equation}
where the exponent $\alpha$ depends on the corresponding quantity from the set $\mathcal{M}$. The expression (\ref{eq_M_1}) correctly reproduces the values of the mechanical properties for the cases considered above -- the monolithic alloy, the nanoporous alloy and the alloy with micron-sized pores, as shown in Figure~\ref{fig_5}. Then, taking into account the conditions (\ref{eq_cond}) and (\ref{eq_conds}) as well as Eqs.~(\ref{eq_Balshin}) and (\ref{eq_M_1}), we find the following general equation
\begin{subequations}\label{eq_model}
\begin{equation}\label{eq_model1}
\mathcal{M}(\bar{l},\,\phi)=\mathcal{M}_{0}\left(\frac{\bar{l}_{0}}{\bar{l}}\right)^{\gamma\phi}(1-\phi)^{n\left[1-\left(\bar{l}_{0}/\bar{l}\right)^{\lambda}\right]}
\end{equation}
on the complete fulfilment of the condition
\begin{equation}\label{eq_model2}
\bar{l}-\bar{l}_{0}\leq\sqrt[3]{\phi V},
\end{equation}
\end{subequations}
which is performed for the full set of mechanical properties $\mathcal{M}=\{E,\,\sigma_{y},\,\sigma_{f}\}$. In Eq. (\ref{eq_model}), the quantity $V$ is the volume of the system.

We are not aware of any models that produce an equation for the mechanical characteristic as a function of two parameters: porosity $\phi$ and average linear pore size $\bar{l}$, as this is accounted for in the proposed equation (\ref{eq_model}). The closest in its physical meaning to our model (\ref{eq_model}) is the Knudsen model with equation
\begin{equation}\label{eq_Knudsen}
\mathcal{M}(G,\,\phi) =\frac{K}{G^{a}}\exp\left(-b\phi\right),
\end{equation}
which is applied to porous ceramics with polycrystalline matrix~\cite{Knudsen_1959}. Equation (\ref{eq_Knudsen}) allows us to approximate the dependence of the strength properties taking into account the porosity $\phi$ and the parameter $G$ characterising the size of crystalline grains in the material. Here, $K$, $a$ and $b$ are empirical parameters. Equation (\ref{eq_Knudsen}) is given as the product of two contributions, where the first contribution is defined as a power-law dependence on the grain size $G$, and the second contribution is defined as an exponentially decaying dependence on the porosity $\phi$. However, it should be noted that the Knudsen model does not apply to porous alloys, but to porous ceramics~\cite{Gu_Zhang_2020,Guevel_2022}.
The Bal'shin model with equation (\ref{eq_Balshin}) does not take into account the dependence of $\mathcal{M}$ on the average linear pore size $\bar{l}$. Nevertheless, it can be generalised to the following form:
\begin{equation}\label{eq_Balshin_gen}
\mathcal{M}(\phi) =\mathcal{M}_{0}(\bar{l})\left(1-\phi\right)^{n(\bar{l})}.
\end{equation}
Comparing (\ref{eq_model}) and (\ref{eq_Balshin_gen}), we can see that the generalization of the Bal'shin model represents only a special case accounted by equation (\ref{eq_model}), where $\mathcal{M}_{0}({\bar{l}_{0}}/{\bar{l}})^{\gamma\phi}$ has a weak dependence on $\phi$.

Equation (\ref{eq_model}) has the following features. First of all, this equation reproduces the nature of the change in mechanical properties with the change in the porosity $\phi$ and in the average linear pore size $\bar{l}$. In this case, the equation depends on the range of values of $\bar{l}$ and $\phi$ defined by relation (\ref{eq_model2}). Here, the quantity $\mathcal{M}_{0}$ is the value of the corresponding mechanical property for the case of the monolithic alloy, and the quantity $\bar{l}_{0}$ is the average interatomic distance, as mentioned above. In the case of monolithic NiTi alloy, we have $\bar{l}_{0}\simeq0.09$\,nm. Equation (\ref{eq_model}) contains three dimensionless parameters $n$, $\gamma$ and $\lambda$ which take positive values. The parameter $n$ is identical to the parameter in the Bal'shin equation, and in the case of NiTi it takes the value $n=3$ for all considered mechanical properties. Furthermore, the parameters $\gamma$ and $\lambda$ characterize how the mechanical property of the material decreases with increasing void size: the larger the value of the parameters $\gamma$ and $\lambda$, the faster the value of the mechanical property decreases. In this case, the parameter $\gamma$ corrects the value of the mechanical characteristic mainly in the limit of large values of the average linear pore size $\bar{l}$. The parameter $\lambda$ is responsible for changing the mechanical characteristic in the limit of small values of $\bar{l}$. It is noteworthy that in the limit of large values of $\bar{l}$, and also when $\gamma\to0$ and $\lambda>>0$, Eq.~(\ref{eq_model}) tends to the Bal'shin equation (\ref{eq_Balshin}). In the case of NiTi, we find that $\gamma=0.028\pm0.002$ and $\lambda=0.25\pm0.03$ for the Young's modulus, while $\gamma=0.65\pm0.04$ and $\lambda= 0$ for the yield strength $\sigma_{y}$ and the ultimate strength $\sigma_{f}$.
\begin{figure*}[ht!]
\centering
	\includegraphics[width=15.0cm]{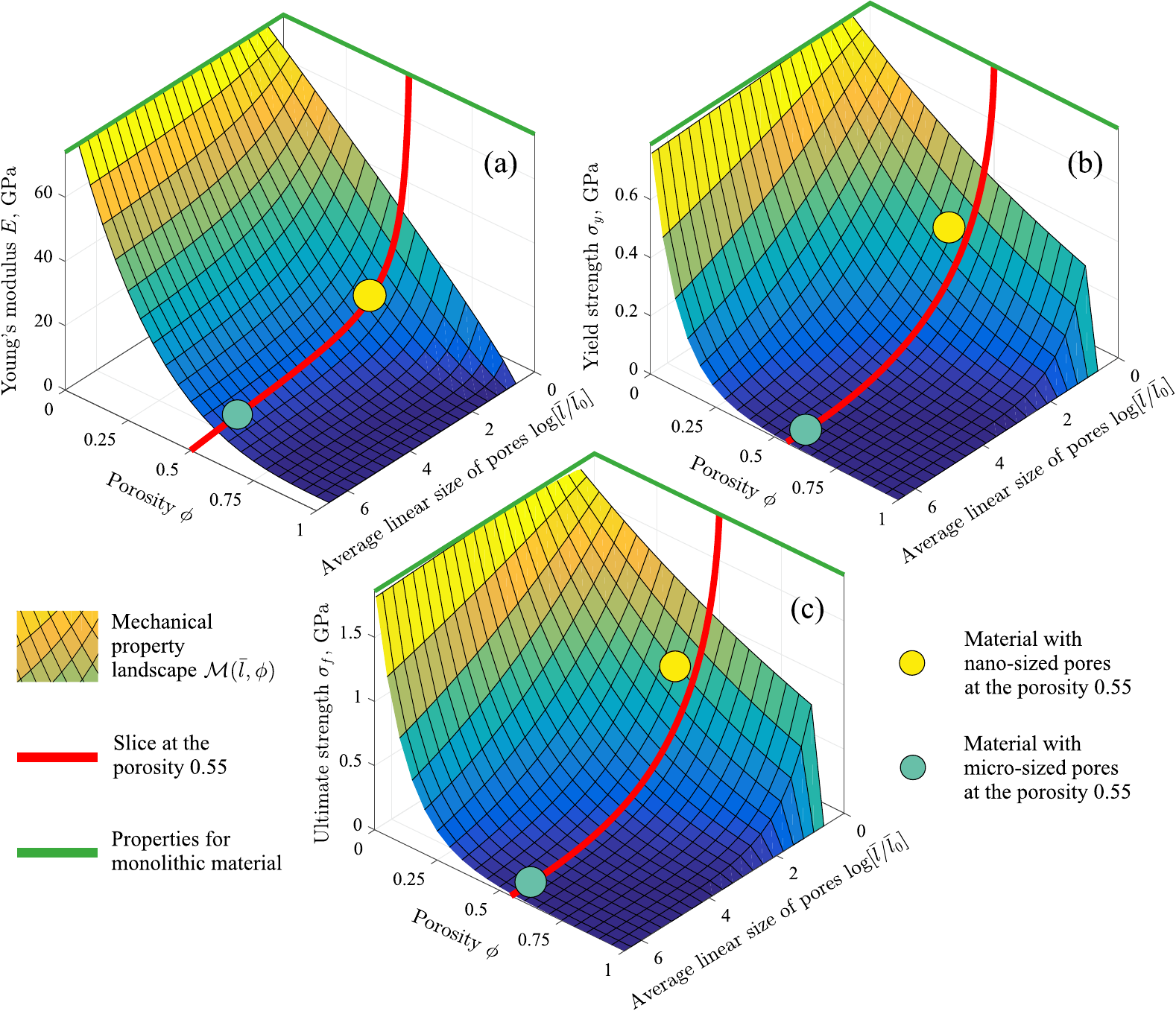}
	\caption{Dependence of the mechanical properties of porous NiTi on the reduced average linear pore size $\log(\bar{l}/\bar{l}_{0})$ and on the porosity $\phi$: (a) Young's modulus $E$, (b) yield strength $\sigma_{y}$ and (c) ultimate strength $\sigma_{f}$. In all these figures, the curved surface is the result of Eq.~(\ref{eq_model}) obtained with different values of $\phi$ and $\bar{l}$. The red curve is the slice of the $\mathcal{M}(\bar{l},\,\phi)$ surface at the porosity $\phi=0.55$. The round markers indicate the mechanical properties found for porous NiTi with nano-sized pores and for NiTi with micron-sized pores.}\label{fig_6}
\end{figure*}

\begin{table}[tbh]
	\caption{Results of Eq.~(\ref{eq_model}) for the mechanical properties of porous NiTi with nano- and micron-sized pores as well as the mean relative error (MRE) $(|\mathcal{M}_{pred}-\mathcal{M}_{calc}|/\mathcal{M}_{calc})\times100\%$, where $\mathcal{M}_{calc}$ are the values of the calculated mechanical properties from Table~\ref{tab_1}, $\mathcal{M}_{pred}$ is the result of Eq.~(\ref{eq_model}).}	
	\begin{tabular}{cccc}
		\hline
		Sample  & $E$, GPa & $\sigma_{y}$, GPa & $\sigma_{f}$, GPa  \\
		\hline
		Nano-sized pores  & $15.9$  & $0.183$ & $0.504$ \\
		Micron-sized pores & $5.989$  & $4.5\times10^{-3}$ & $17.9\times10^{-3}$ \\
		MRE, \% & $\approx4.3$ & $\approx28$ & $\approx28$ \\
		\hline
	\end{tabular}\label{tab_2}
\end{table}

Figure~\ref{fig_6} shows the Young's modulus $E$, the yield stress $\sigma_{y}$ and the ultimate strength $\sigma_{f}$ as functions of the porosity $\phi$ and the average linear pore size $\bar{l}$ calculated for NiTi alloy using Eq.~(\ref{eq_model}). It can be seen from this figure that the values of the mechanical properties approach the strength parameters of monolithic NiTi in the limit of small $\phi$ and $\bar{l}$. On the other hand, when the pore sizes become comparable to the size of the system, i.e. $\bar{l}-\bar{l}_{0}=\sqrt[3]{\phi V}$, the theoretical limit is reached, which satisfies the condition (\ref{eq_philimit}), under which the material does not physically exist. At the same time, the change in the mechanical properties is more pronounced at the porosity $\phi\in(0,\,0.6]$ and/or at the average linear pore sizes $\bar{l}<10$~nm. The same result is obtained from the analysis of  Eq.~(\ref{eq_model}) with the parameters $n$, $\gamma$ and $\lambda$ defined for NiTi. Thus, the introduction of insignificant porosity into an almost homogeneous monolithic material leads to a significant reduction in its mechanical properties. In the limit of high porosity (i.e. at $\phi>0.6$) and the average linear pore size corresponding to a highly porous material with  micron or millimeter sized pores, the mechanical properties are weakly dependent on $\phi$ and $\bar{l}$. Such porosity conditions correspond to a material in which the porous structure is percolating, and the increase in porosity is mainly due to the thinning of the interpore partitions.

The mean relative error (MRE) between the calculated values of the mechanical properties and the result of Eq.~(\ref{eq_model}) is $\approx4.3$\,\% in the case of the Young's modulus $E$ [see Table~\ref{tab_2}]. For the yield strength $\sigma_{y}$ and the ultimate strength $\sigma_{f}$, the MRE takes the value $\sim28$\,\%, which is to be expected since the values of these quantities can vary over a wide range~\cite{S_Hellberg_2020,A_Jalaeefar_2013,Metals_2023}. In this case, the relative error can be reduced by analyzing a large set of empirical and simulation data for NiTi samples with different porosity and with different average linear pore size. In addition, given the recent advances in the application of the artificial intelligence methodology to solving physical problems, it is expected that the problem of increasing the amount of data for the mechanical properties can be solved using machine learning methods~\cite{Galimzyanov_Materials2023,Galimzyanov_PhysicaA2023}.

\section{Conclusions}

Thus, in the present work, a comprehensive experimental and simulation study of the fracture process of porous crystalline NiTi under uniaxial stress has been investigated. In the experimental part of the work, porous NiTi with porosity $\phi=0.55$ and micron-sized pores were synthesized. By structural analysis of the porous samples, the distribution of pores by linear size was obtained and the average linear pore size $\bar{l}$ was estimated. In addition, the mechanical properties of the porous NiTi samples were calculated in tensile tests. The statistical treatment of the obtained empirical data allowed to determine the most probable stress-strain curve and to correctly estimate the Young's modulus, the yield strength and the ultimate tensile strength. In the simulation part, the fracture process of porous crystalline NiTi under tensile stress was considered by the method of non-equilibrium molecular dynamics simulation. For the microscopic scales, which are related to the scales on which the interparticle interaction potential acts, the information about the mechanical properties has been obtained. The porous system has the porosity $\phi=0.55$, and this system consists of nano-sized pores. For this system, the most probable stress-strain diagram is determined by averaging the results obtained from different independent molecular dynamic simulations. In addition, the general methodology for determining key characteristics of porous materials such as pore size distribution, interpore partition size distribution, average linear pore size, average partition thickness and porosity is discussed. It is shown that these characteristics are related to the mechanical properties within the framework of well-known models such as the Bal'shin and Ryshkewitch-Duckworth equations.

The analysis of the obtained results showed that the average linear pore size $\bar{l}$ is an important factor in addition to the porosity $\phi$ in the estimation of the mechanical properties of porous materials. A condition is formulated which shows that a porous material can be realized only in a strictly defined range of values of $\phi$ and $\bar{l}$. The results of the present work have been supplemented with data for the case of a monolithic NiTi alloy, which allowed us to obtain a completely new general equation (\ref{eq_model}), which defines the dependence of an arbitrary mechanical characteristic on the average linear pore size and on the porosity. Equation (\ref{eq_model}) correctly considers the limiting case corresponding to a monolithic material (i.e., at the porosity $\phi\rightarrow0$) as well as the case of no material (i.e., at the porosity $\phi\rightarrow1$). Using porous NiTi as an example, it was shown that Eq.~(\ref{eq_model}) correctly reproduces the dependence of the Young's modulus, the yield strength and the ultimate tensile strength of porous NiTi on the average linear pore size. This equation can be adapted to predict the strength properties of porous alloys with different compositions and porosity parameters if sufficient empirical and simulation data are available.

The results obtained in the present study allow us to approach the solution of actual problems of fundamental character, which may include: (i) understanding the mechanisms by which crack formation and growth occur in porous metal alloys; (ii) understanding how porous structure parameters, such as pore size distribution, interpore partition size distribution, pore shape, pore wall surface area, porosity, etc., affect the elastic and plastic properties and strength of porous metal alloys; (iii) explanation of the influence of structural heterogeneity arising due to the non-sphericity of pores, their orientation and distribution in the system on the mechanical properties of the porous alloy. The solution of these problems is only possible by combining experimental results for macroscopic mechanical properties with the results of molecular dynamics simulations of non-equilibrium processes on microscopic scales.

\section*{Acknowledgment}
The work was supported by the Russian Science Foundation (project No. 19-12-00022, \\ https://rscf.ru/project/19-12-00022/, accessed on 06 December 2023). S.G.A. and N.V.A. are grateful to the Kazan Federal University Strategic Academic Leadership Program (PRIORITY-2030), which supported the experimental part of the work.

\section*{Appendix: Experimental and simulation details}

\subsection*{Experimental synthesis of porous NiTi samples}

Porous NiTi samples were obtained by single sintering using NiTi powders of grade PV-N55T45 (manufacturer: Polema, Russia) with the fraction $100$-$200$~$\mu$m, which have a two-phase state: B2-austenite ($30$~wt.~\%) and B19'-martensite ($34$~wt.~\%)~\cite{Anikeev_Garin_2018,Anikeev_Artyukhova_2022}. The powders contain phases enriched in titanium Ti$_2$Ni ($26$~wt.~\%) and nickel -- TiNi$_3$ ($3$~wt.~\%), as well as metastable phases Ti$_3$Ni$_4$ ($7$~wt.~\%). Single sintering was carried out in the electric vacuum furnace SNVE-1.31/16-I4 (manufacturer: VARP Vacuum, Russia) for $15$~min at the temperature $1528\pm5$\,K. The temperature was controlled by means of a tungsten-rhenium alloy thermocouple. The specified temperature-time regime allows one to obtain experimental samples with an optimal degree of sintering with a high quality of interparticle contacts, a regular porous structure and the lowest possible degree of shrinkage. Sintering was carried out at the pressure $6.65\cdot10^{-4}$\,Pa with the average heating rate $10$\,K/min. The porosity of the samples was determined by the weighing method and is equal to $\phi=0.55$. The porous samples have a cylindrical shape with the diameter $11$--$13$\,mm and the length $65$--$80$\,mm. The sizes of the pores and the interpore partitions were determined by combination of the cutting plane method. The study of physical and mechanical properties was carried out by the tensile method on the universal electromechanical testing machine Instron 3369 (manufacturer: Instron, USA) at a strain rate not higher than $7.5\times10^{-6}$\,m/sec. The surface of the pore walls, the microstructure and the fracture fractograms of the samples were studied by using Quanta 3D scanning electron microscopes (manufacturer: FEI Company, USA) in the secondary electron mode in high vacuum and at the accelerating voltages from $20$ to $30$\,kV and the beam sizes $5$--$20$\,nm.

\subsection*{Computational details and methods}

Molecular dynamics simulation of crystalline NiTi has been carried out for the sample with the B2 crystal lattice consisting $12\,577$ atoms of Ni and $12\,577$ atoms of Ti. The atoms have been placed in a cubic system with the linear size $L\simeq9$~nm. The interaction between Ni and Ti atoms is determined using the second nearest neighbour modified embedded atom method (2NN MEAM)~\cite{Tsygankov_Galimzyanov_Mokshin_2022,Lee_Baskes_2000,Ko_2015}:
\begin{equation}\label{eq_2nnmeam_pot}
E = \sum_{i=1}^{N}\left[F_{i}(\rho_{i})+\frac{1}{2}\sum_{j\neq i}^{N}S_{ij}\phi_{ij}(r_{ij})\right].
\end{equation}
This potential reproduces the physical properties of NiTi for a wide thermodynamic region, as has been shown previously in the works~\cite{Jiavi_Chen_2021,Jeongwoo_Lee_2021,Yang_Guo_2017}. In Eq.~(\ref{eq_2nnmeam_pot}), $E$ is the total energy of the system; $F_i(\rho_i)$ is the embedding function for the atom $i$ within the background electron density $\rho_i$; the pair potential $\phi_{ij}(r_{ij})$ and the screening function $S_{ij}$ are evaluated at the distance $r_{ij}$ between atoms $i$ and $j$; the cutoff radius of this potential is $5.0$~\AA~for the considered binary system.

Porous NiTi samples were formed by cutting out the voids through by removing some of the atoms from the monolithic system. Only those atoms that were located inside imaginary ellipsoids were removed from the system. The positions and spatial orientations of these ellipsoids were determined randomly. At the same time, the ellipsoids can intersect with each other to form open voids. The values of the semi-axes of the ellipsoids were set randomly and were chosen from the range $[2.0;\,6.0]\,$nm. Changing the semi-axes of the ellipsoids in the considered range makes it possible to obtain an open-type porous structure with the porosity $\phi=0.55$, which coincides with the porosity of the experimental samples. After cutting out the voids, the sample was brought to a state of thermodynamic equilibrium at the temperature $T=300\,$K and the pressure $1$\,atm for the time $60$\,ps. This time is sufficient to minimise the free energy of the void surfaces and to obtain a stable porous structure. The molecular dynamics simulation was performed using the Large-scale Atomic/Molecular Massively Parallel Simulator (LAMMPS) software (Sandia National Laboratories, USA)~\cite{Thompson_Aktulga_2022}. The 3D visualisation of the simulation results and the analysis of the porous structure were performed using the OVITO software (OVITO GmbH, Germany)~\cite{Stukowski_2010}.

The resulting porous samples were subjected to uniaxial tensile stress along the OX axis at the constant rate $1.0\cdot10^{10}$\,s$^{-1}$. It is important to note that such ultra-high strain rates are typical for the molecular dynamics simulation method used to study the process of material destruction under high dynamic loading~\cite{J_Li_2019,Q_Zeng_2021}. This rate differs from the experimental strain rate, which can lead to overestimation of mechanical properties in the case of molecular dynamics simulation, for example, as shown in Refs.~\cite{J_Herman_2020,Piao_Huh_2016}. The value of the stress $\sigma_{xx}$ was determined using the Irwin-Kirkwood equation~\cite{Tsai_D_1979}:
\begin{equation}\label{eq_IK}
\sigma_{xx}=-\frac{1}{V}\left(\sum_{i=1}^{N}m_{i}\vartheta_{xi}^{2}+\sum_{i=1}^{N}\sum_{j>i}^{N}r_{ijx}F_{ijx}\right).
\end{equation}
Here, $V$ is the volume of the system; $m_{i}$ is the mass of the $i$th atom; $\vartheta_{xi}$ is the $x$ velocity component of the $i$th atom; $F_{ijx}$ is the $x$ component of the force between the particles with the labels $i$ and $j$; $r_{ijx}$ is the $x$ component of the distance between the particles $i$ and $j$. The tensile strain $\epsilon$ was determined by the relation $\epsilon(t)=(L_{x}(t)-L_{x}^{(init)})/L_{x}^{(init)}$, where, $L_{x}^{(init)}$ is the length of the simulation box before deformation, and $L_{x}(t)$ is this length at time $t$ after the start of deformation.

\end{document}